\documentclass[twocolumn]{article}
\PassOptionsToPackage{dvipsnames,table}{xcolor} % Some package in `preprint` (probably `background`) includes `xcolor`.
\usepackage{preprint}

\SetBgContents{}

\usepackage[utf8]{inputenc}
\usepackage[T1]{fontenc}
\usepackage[english]{babel}
\usepackage{csquotes}
\usepackage[final,
            tracking=true,
            kerning=true]{microtype}
\usepackage{xspace}

\usepackage{amssymb}
\usepackage{amsmath}
\usepackage{amsthm}
\usepackage{mathtools}
\usepackage{bbm}

\usepackage{graphicx}
\usepackage{booktabs}
\usepackage{subcaption}
\usepackage{listings}
\usepackage{float}

\usepackage{nth}
\usepackage{siunitx}
\usepackage{color}
\usepackage[normalem]{ulem}

\newcommand*{\titletext}{Intuitiveness in Active Teaching}

\usepackage[backend=biber,
            sorting=nyt,
            maxcitenames=2,
            maxbibnames=8,
            backref=true]{biblatex}
\usepackage{varioref}
\usepackage[hypertexnames=false,
            colorlinks=true,
            linkcolor=MidnightBlue,
            urlcolor=Maroon,
            citecolor=ForestGreen,
            pdfauthor={Jan Philip Göpfert},
            pdftitle={\titletext}]{hyperref}
\usepackage[all]{hypcap}
\usepackage[noabbrev,
            capitalize,
            nameinlink]{cleveref}
\crefname{hypothesis}{Hypothesis}{Hypotheses}

\newfloat{lstfloat}{htbp}{lop}
\floatname{lstfloat}{Listing}

\crefalias{lstfloat}{listing}

\newcommand*{\eg}{e.\,g.\@\xspace}
\newcommand*{\ie}{i.\,e.\@\xspace}
\newcommand*{\cf}{cf.\@\xspace}

\newcommand*{\vs}{vs.\@\xspace}
\newcommand*{\R}{\mathbb{R}}

\DeclarePairedDelimiterX{\norm}[1]{\lVert}{\rVert}{#1}

\newcommand*{\X}{\mathcal{X}}  % feature space
\newcommand*{\Y}{\mathcal{Y}}  % label space
\newcommand*{\nnwith}{\texttt{NN-with}\@\xspace}
\newcommand*{\nnwithout}{\texttt{NN-without}\@\xspace}
\newcommand*{\dtwith}{\texttt{DT-with}\@\xspace}
\newcommand*{\dtwithout}{\texttt{DT-without}\@\xspace}

\newcommand*{\p}{p}

\newcommand*{\worldmapwidth}{3.5cm}

\title{\titletext}

\usepackage[auth-sc]{authblk}

\author[1]{Jan Philip Göpfert}
\author[1]{Ulrike Kuhl}
\author[1]{Lukas Hindemith}
\author[2]{Heiko Wersing}
\author[1]{Barbara Hammer}

\affil[1]{Bielefeld University, Germany}
\affil[2]{Honda Research Institute Europe, Offenbach, Germany}

\begin{document}

\twocolumn[
\begin{@twocolumnfalse}
\maketitle
\begin{abstract}
While Machine learning gives rise to astonishing results in automated systems, 
it is usually at the cost of large data requirements.
This makes many successful algorithms from machine learning unsuitable for human-machine interaction,
where the machine must learn from a small number of training samples that can be provided by a user within a reasonable time frame.
Fortunately, the user can tailor the training data they create to be as useful as possible,
severely limiting its necessary size -- as long as they know about the machine's requirements and limitations.
Of course, acquiring this knowledge can in turn be cumbersome and costly.
This raises the question of how easy machine learning algorithms are to interact with.
In this work, we address this issue by analyzing the intuitiveness of certain algorithms when they are actively taught by users.
After developing a theoretical framework of intuitiveness as a property of algorithms, 
we introduce an active teaching paradigm involving a prototypical two-dimensional spatial learning task as a method to judge the efficacy of human-machine interactions.
Finally, we present and discuss the results of a large-scale user study into the performance and teaching strategies of \num{800}~users interacting with two prominent machine learning algorithms in our system,
providing first evidence for the role of intuition as an important factor impacting human-machine interaction.
\end{abstract}
\vspace{0.5cm}
\end{@twocolumnfalse}
]

\section{Introduction}\label{sec:introduction}

\subsection{Problem Setting}\label{subsec:problem-setting}
Human-machine interaction can be overt, \eg, when a person engages in spoken dialog with a humanoid robot;
or it can be obscured, \eg, when they formulate a natural-language query to enter into a search engine on the Internet.

Pioneering work in the field of human-machine interaction has shown that when a \emph{naive user} interacts with an artificial system, they react to them as they would to another person, following rules of social conduct~\autocite{Nass2000MachinesMindlessness,Nass1996ComputersTeammates}: “Computers are Social Actors.”~\autocite{Nass1994ComputersSocialActors}
Users automatically and subconsciously neglect artificial characteristics of the machine and follow certain expectations and rules drawn from prior human-human interaction~\autocite{Nass2000MachinesMindlessness,Nass1996ComputersTeammates,Nass1994ComputersSocialActors}.
Thus, just as in interpersonal communication, any user enters human-machine interactions with inherent assumptions and quickly forms a \emph{cognitive representation} of the entity they interact with. This representation may change dynamically over the course of the interaction.

When the interaction has a measurable goal (\eg, the user and robot successfully carrying out a task, or the user finding an answer to a question among the search engine's responses), we can relate this measure to the usefulness of the cognitive representation.

Much of the recent success within artificial intelligence can be attributed to advances in machine learning (ML), especially supervised learning, where a predictor (called “model”) \(f \colon \X \to \Y\) is learned by the machine based on a training set \((X, Y) \subset \X \times \Y\).
The efficacy of \(f\) depends largely on the algorithm used for training, the quality of the available training set \((X, Y)\), and the difficulty of the underlying problem, \ie, the true relationship between \(\X\) and \(\Y\).

Usually, first a training set is gathered, then a model \(f\) is trained, and finally, this trained model is used for prediction.
However, when circumstances change and the initial training set no longer viably represents the underlying problem, the process needs to be repeated and a new model \(f'\) replaces the old one.
Because circumstances in the real world perpetually change, a lot of research revolves around the development of adaptive systems that learn incrementally and gracefully handle new data as well as changes in the underlying distribution~\autocite{Parisi2019LifelongLearning,Widmer1996LearningConceptDrift}.
This naturally entails a number of problems, mostly related to costs (monetary or otherwise) of incrementally acquiring data, training a model, and the model's deployment.

The phenomenon of a ML model needing to adapt to new data is perhaps most pronounced when it is supposed to learn and apply acquired knowledge during a real-world interaction with a human user.
In the future -- at least within the European Union -- artificial intelligence will be legally required to adapt to new data~\autocite{EU2020AI}.
While there are other approaches to dealing with limited or changing data (\eg, memory-augmented one-shot learning\autocite{Vinyals2016OneShot}),
how to efficiently model, train, and deploy such systems is still a matter of ongoing inquiries and its own research avenue~\autocite{Paaßen2020ReservoirMemory}.

\subsection{Active Learning}\label{subsec:active-learning}
In traditional batch learning, training data is obtained independently of the later training.
This does not necessarily need to be the case.
For instance, in active learning~\autocite{Settles2009ActiveLearning}, a model is trained incrementally:
after learning from a training set \((X_i, Y_i)\), the model \(f_i\) self-evaluates and determines another input set \(\X_{i + 1}\) for which it requests labels \(Y_{i + 1}\), whereby it extends the training data, which is in turn used to update the model \(f_{i+1}\).
In principle, this process can be repeated indefinitely.
Typically, the goal is to keep the number of training points small while achieving high performance.

While the model can request labels for arbitrary points in \(\X\) from a so-called \emph{oracle}, it must have some sort of strategy for how to choose the most useful points for labeling.
When the model includes a notion of certainty about its prediction at a given point, this can be used to perform various types of \emph{uncertainty sampling}~\autocite{Lewis1994SequentialAlgorithm,Budd2019SurveyActiveLearning}, where the model requests a label for the point about which it is least certain.

\subsection{Machine Teaching}\label{subsec:machine-teaching}
While in active learning it is the learning algorithm that chooses points to get labeled by an oracle, this role can also be assumed by the oracle: in machine teaching~
\autocite{Zhu2015MachineTeachingInverse,Simard2017MachineTeaching,Zhu2018OverviewMachineTeaching}, one searches a small training set with which a given learning algorithm performs best.

The teaching dimension is a theoretical concept that characterizes the amount of data necessary to \emph{teach} a concept.
The \emph{recursive} teaching dimension relies on the further assumption that the learner knows that they are being taught -- by contrast, the Vapnik Chervonenkis (VC) dimension characterizes the amount of randomly distributed data required to \emph{learn} a concept~\autocite{Goldman1995TeachingComplexity,Kobayashi2009TeachingComplexity}.
When the teacher knows about the teaching problem, including the underlying truth and the inner workings of the learning algorithm, they pick samples deliberately instead of randomly.
For certain cases, the (recursive) teaching dimension can be upper bounded by the VC dimension,
but the two quantities can be arbitrarily far apart~\autocite{BenDavid1998SelfDirected,Doliwa2014Recursive}.

\subsection{Active Teaching}\label{subsec:active-teaching}
In \emph{iterative machine teaching}, an informed teacher feeds examples iteratively to a learner based on the current performance of the learner~\autocite{Mathias1997InteractiveTeaching, Liu2017IterativeMachineTeaching}.
Omniscient teachers have been observed to perform exponentially better than random ones in certain cases~\autocite{Liu2017IterativeMachineTeaching,Barz2019QuestionAnswering}, despite there being no known optimal teaching algorithm for a large number of practical learning problems~\autocite{Cakmak2014GoodTeaching}.
Mathematical considerations rely on an assumed rationality of learner and teacher, and human teaching behavior can deviate from an optimal one~\autocite{Cakmak2014GoodTeaching}.

In our work, we explore the question of how naive users act in a teaching task within a natural two-dimensional geometric setting,
whether their ability to teach can improve based on natural interaction with a ML algorithm,
and whether there exist differences between different learners.

Consider an actual real-world interaction between a user and a machine, \eg, someone telling a service robot about what to call different areas in their home and where those areas are.
In this specific case, the objective is for the machine's layout of the home to match the user's (which we consider the truth) and to achieve this with a small number of training samples.
Because the user has perfect knowledge of the underlying truth
(insofar as they know the arrangement of the rooms and they themselves decide what a given region should be referred to),
this situation is similar to machine teaching.
However, the user does not necessarily know about the algorithms at work inside the robot.
Instead, since this is a true interaction, they can teach the robot incrementally and observe the updated state after each exchange.

We think of this setting as \emph{active teaching}, because in it, the teacher actively selects training samples to efficiently teach a machine, while using immediate feedback to develop an understanding of what constitutes a useful training sample.
In essence, the teacher infers information about the learner based on how their provided samples are processed.
The teacher is not restricted to passively observing the learner, nor does the teacher need to start out with perfect knowledge about the learner.
In other words, active teaching entails a reciprocal learning process: While the machine learns based on the data provided by the user, the user learns, with each iteration, how to select more effective training samples.

Ultimately, understanding the dynamics of active teaching this will lead to improved performance of more reliable machines~\autocite{Duros2019Intuition}.
Because we want such intelligent systems to be able to work with the general population~\autocite{Simard2017MachineTeaching}, we must ask ourselves how naive users approach cooperation with a machine without explicit prior training.

\subsection{Intuitiveness in Active Teaching}\label{subsec:intuitiveness-in-active-teaching}
To make machines approachable and easily usable, we aspire to human-machine interfaces that do not require extensive studies of manuals or prior training.
Intuitive problem solving has been compared against systematic problem solving in manufacturing~\autocite{Tyre1993SystematicIntuitive}, with a focus on problem-solving strategies.
In such a setting, systematic approaches are superior.
This, however, does not diminish the importance of intuition in human-machine interaction.
Considerable efforts have been conducted towards designing technical systems that take into account non-conscious use of previous knowledge~\autocite{Naumann2007IntuitiveUserInterfaces}.

While the ubiquitous nature of technology is without question, to date there has been little research into how easily accessible ML algorithms are to lay users -- even though ML has become an integral part of modern information technology.
In interactive scenarios with machines that employ ML, we would like those machines to \emph{just work}.
Thus, we must ask: “Are ML algorithms intuitive?”
To answer that question, we consider the notion of intuitiveness as a property of algorithms.

In \cref{sec:experimental-design} we introduce a problem that a user must solve cooperatively by teaching a machine, without being provided any explanations or inside knowledge about the machine.
This way, we leave them only with their intuition to learn from the interactions they themselves control.

\section{Theoretical Foundation}\label{sec:theoretical-foundation}

\subsection{Intuitiveness}\label{subsec:intuitiveness}
In everyday life, one often talks about \emph{gut feelings} or \emph{hunches} when reaching decisions intuitively instead of analytically.
However, what exactly constitutes \emph{intuitive problem solving and decision making} is still a matter of debate.
While exact definitions of intuition vary in the literature, a common theme is to describe it as an instantaneous, automatic, experience-driven, and unconscious way of information processing~\autocite{Epstein1996IndividualDifferences,Glockner2008MultipleReason,Bowers1990Intuition}, often contrasted with slow and effortful weighing of options and facts underlying more analytical thought processes.
In accordance with this dichotomy, there is a growing line of research that considers two distinct systems of processing that are active in the human brain during problem solving~\autocite{Epstein1994CognitivePsychodynamic,Kahneman2011ThinkingFastSlow}:
 \emph{System~I}, fast, effortless, and difficult to control;
and \emph{System~II}, comparatively slow, effortful, and consciously more accessible~\autocite{Kahneman2003JudgmentChoice}.

Just as difficult as finding a concise definition of intuitive decision making itself, is characterizing what makes an object or task intuitively understandable.
However, such a characterization is vital in the type of human-machine interaction explained in \cref{sec:introduction}, as it can facilitate deeper insights into what ML approaches users may easily interact with, even without previous training.
With regards to technical applications, the term \emph{intuitiveness} is used to refer to the design facilitating situation-specific operations,
\eg, information retrieval~~\autocite{Zhou2013Intuitiveness}, quantity control~\autocite{Lylykangas2013Intuitiveness}, and nonverbal communication~\autocite{Grandhi2011Intuitiveness}.

\emph{We say an algorithm is \emph{intuitive}, if a user interacting collaboratively with the algorithm gets better over time with regards to the learning objective, without receiving explicit training or explanations before or during the interaction; within an overall brief time frame.} 

Let us contemplate the individual pieces of this definition, considering classification algorithms.

\subsubsection{Collaboration \& Shared Learning Objective}\label{subsubsec:collaboration}
As outlined above, in our active teaching paradigm the user supplies training samples of the form \((x, y) \in \X \times \Y\) to the algorithm, which are used to form a training set with which to learn a function \(f \colon \X \to \Y\).
Given a further test set \((X_\text{test}, Y_\text{test}) \in \X \times \Y\), we can measure the accuracy of \(f\) as the number of correct predictions divided by the total number of predictions \(\lvert X_\text{test} \rvert\).
As long as the algorithm to train \(f\) and the underlying true relation between \(\X\) and \(\Y\) remain unchanged and deterministic, the accuracy is determined only by the user's selection of training samples.
Therefore, we can directly link the model's accuracy to the performance of the user: the user and the algorithm share a common objective.

After a new training sample is added to the training set, the model is updated accordingly.
Communicating the updated model to the user makes the collaboration truly interactive.
This way, they can draw conclusions about how their choice of training samples affects the resulting model.

\subsubsection{Improvement}
If we know the smallest number of training samples required (\cf \cref{subsec:machine-teaching}) or can at least bound it, we can expect the accuracy of the trained model to increase while the number of training samples approaches that number, provided the user chooses adequate training samples.

Given our definition of intuitiveness above, a user interacting with an intuitive algorithm will consistently improve in their choice of training samples.
This we can observe when the same user is presented with a new problem (in the form of a different relation between \(\X\) and \(\Y\)) which they solve collaboratively with the same algorithm as before.

\subsubsection{Lack of Explanations}\label{subsec:lack-of-explanations}
If the user receives explanations on how to solve the problem or useful information about the algorithm, they no longer need to rely only on their intuition to reach useful conclusions about their interaction.
Therefore, when explicitly evaluating how easily accessible or intuitive the respective algorithm is, any such external information must be kept hidden from the user.

Of course, the user still needs to understand the objective they share with the algorithm, and they need some way of inspecting the current state of the trained model in order to successfully build an understanding of the algorithm and their interaction.

\subsubsection{Limited Time}

What constitutes a \emph{brief time frame} is subjective and depends on the circumstances.
With regards to \emph{intuitiveness}, it is important to create a setting in which the user remains free of external influence.
We achieve this by keeping the interactive collaboration uninterrupted;
this implies a single, continuous session, which in turn limits the overall time that is available.

\subsection{Hypotheses}\label{subsec:hypotheses}
Equipped with the definition from \cref{subsec:intuitiveness},
we formulate the following hypotheses about intuitiveness of ML algorithms,
and we will test them in the remainder of this work:
\begin{enumerate}
    \item\label[hypothesis]{hypothesis1} There are intuitive ML algorithms.
    \item\label[hypothesis]{hypothesis2} Different ML algorithms differ with respect to how intuitive they are.
    \item\label[hypothesis]{hypothesis3} Even without external explanations, different visualizations of the current state influence how quickly a user forms a useful cognitive representation of the algorithm they interact with.
\end{enumerate}

To investigate these hypotheses, we designed an active teaching study in which naive users taught an algorithm about the layout of different areas on a two-dimensional image by selecting only a limited amount of training pixels on the image.
This human-grounded evaluation~\autocite{Doshi2017RigorousInterpretable} leads us to new insights into general behavioral patterns of naive users.

\subsection{Intuitiveness as a Property of Algorithms}\label{subsec:intuitiveness-algorithms}
Before we lay out the experimental setup of our study,
let us revisit our definition of intuitive algorithms in light of the hypotheses we intend to test.

Assume an intuitive algorithm exists; to actually observe its intuitiveness,
we need to witness a user improving their interaction with the algorithm.
During a suitable experiment, all other relevant factors need to allow for intuitive learning and an increase in performance as well:
\begin{itemize}
    \item Although limited, the available time must suffice for the user to learn, and for the algorithm to show improved performance.
    \item Even without receiving explicit information, the user needs to be able to process the feedback they receive from the learning algorithm.
    \item The set task must be easy enough to be tackled by an algorithm taught by a teacher who relies on their intuition, but challenging enough for notable improvement to occur.
\end{itemize}
If we failed to carefully design the experiment accordingly,
we might not be able to observe the algorithm's intuitiveness.
On the other hand, if we do observe an improvement in performance as required by our definition,
we can indeed attribute intuitiveness to the algorithm,
as well as recognize that all other aspects of the procedure allow for intuitiveness.

\section{Experimental Design}\label{sec:experimental-design}

To test \cref{hypothesis1,hypothesis2,hypothesis3} from \cref{subsec:hypotheses} we selected two algorithms that satisfy a number of criteria:
\begin{itemize}
    \item They should be well established in ML and relevant.
    \item A small number of training samples must be sufficient in order to keep the experiment short.
    Furthermore, it should be possible for a single new training sample to make a visible difference.
    \item The algorithms cannot exhibit overly complex behavior.
\end{itemize}
The ML library \emph{scikit-learn}\footnote{\url{https://scikit-learn.org/}} provides implementations of \emph{nearest neighbors} classification and \emph{decision tree} classification, which perfectly fit our requirements
\footnote{
Before settling on this choice of algorithms,
we experimented with a vast number of additional algorithms and hyperparameters,
and we observed how users would interact with them.
We learned that more complex algorithms,
especially those with nonlinear decision boundaries and regularization,
required substantially more training points and larger world maps (see \cref{subsec:ground-truth-generation-visualization}) to allow a teaching process that felt truly interactive.
At the same time, even the simplest algorithms proved initially challenging for users unfamiliar with ML,
and their reasonable computational requirements ease real-time interaction for multiple concurrent users on relatively simple hardware.
}.
We briefly introduce these algorithms in \cref{subsec:nearest-neighbors,subsec:decision-trees}.
The shared goal in our active teaching scenario is to learn the boundaries between regions in images, which has several advantageous properties:
\begin{itemize}
    \item We can expect our participants to have at least some familiarity with maps, specifically political maps that use color to visualize boundaries between regions.
    \item Learning, and by extension teaching, is directly related to cognitive maps~\autocite{Tolman1948CognitiveMaps}.
    \item Teaching arbitrary labels (in our case colors) is directly useful in a number of applications.
    \item Embedding complex data in two dimensions for visualization and interaction is ubiquitous in technical applications across a number of professional domains,
    underlining the broad applicability of our results~\autocite{Madl2015CognitiveSpatialMemory}.
\end{itemize}
See \cref{subsec:ground-truth-generation-visualization,subsec:concurrent-visualization-of-ground-truth-and-taught-model,subsec:teaching} for details on how we create map-like images and how training samples are communicated.

\subsection{Nearest Neighbors}\label{subsec:nearest-neighbors}
The prime example of nonparametric methods, nearest neighbors algorithms work directly on the training data during prediction.
Given a training set \((X, Y) \in \X \times \Y\) and a query point \(x \in \X\),
the \(k\)~points in \(X\) nearest to \(x\) are calculated and the most frequent of their labels is assigned to \(x\).
Setting \(k = 1\) effectively segments \(\X\) into uniformly labeled Voronoi cells, with training points at their centers.
In two dimensions, these Voronoi cells can be readily visualized.

\subsection{Decision Trees}\label{subsec:decision-trees}
Decision trees are collections of if-then-else rules.
Given an input \(x \in \X\),
the outcome of a rule determines which rule to apply next, and so on, until a rule is reached that assigns a label to \(x\).

A decision tree is constructed from a given training set by consecutively splitting it, according to the specific learning algorithm~\autocite{Shalev2014UnderstandingMachineLearning} -- in our case CART~\autocite{Breiman1984CART} with Gini Impurity.

\subsection{Ground Truth Generation \& Visualization}\label{subsec:ground-truth-generation-visualization}
\begin{figure}
    \centering
    \subcaptionbox{Nearest Neighbor\label{subfig:ground-truth-nn}}{
        \includegraphics[width=\worldmapwidth]{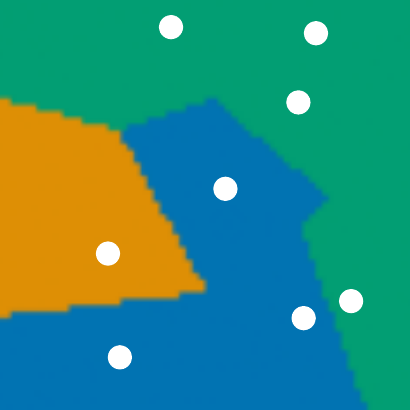}
    }
    \subcaptionbox{Decision Tree\label{subfig:ground-truth-tree}}{
        \includegraphics[width=\worldmapwidth]{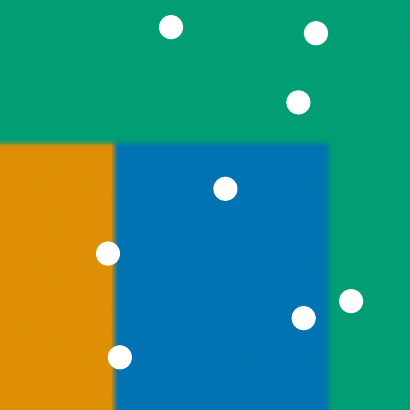}
    }
    \caption{World maps using a nearest neighbor classifier and a decision tree classifier, respectively.
    We indicate positions of the generating points underlying this world map with white circles -- these are never shown to users.
    }
    \label{fig:ground-truth}
\end{figure}

Participants of our study teach (see \cref{subsec:teaching}) regions of a two-dimensional image to either the nearest neighbor or the decision tree algorithm.
We refer to these \num{64}-by-\num{64} images as \emph{world maps}.
To generate one for a given classification algorithm,
we sample and label a set number of points randomly within the (initially empty) world map
and use this \emph{generating set} to train a \emph{generating classifier}.
We use this classifier to “predict” the label of each pixel and color\footnote{We use the \texttt{colorblind} palette provided by \emph{seaborn}: \url{https://seaborn.pydata.org}} it accordingly (see \cref{fig:ground-truth}).

Through this process we can create and visualize a practically infinite number of ground-truth relations between \(\X = \R^2\) and \(\Y = \{1, \dots, m\}\), where \(m\) is the number of classes.
It also ensures that the taught algorithm can in fact reach a perfect solution if the generating set is provided for training.

We fix the number of classes \(m = 3\) and the number of generating points \(n = 8\) for each world map.
To preclude too simplistic world maps we discard any world map that does not have at least one out of six pixels assigned to each color
or in which two generating points are extremely close to each other.

We generate \num{13}~fixed world maps for each classification algorithm,
and randomly pick one that we show to every user in the beginning.
The remaining \num{12}~world maps are presented in an order randomized for each user.

\subsection{Concurrent Visualization of Ground Truth and Learner}\label{subsec:concurrent-visualization-of-ground-truth-and-taught-model}
\begin{figure}
    \centering
    \subcaptionbox{Nearest Neighbor\label{subfig:overlayed-boundaries-nn}}{
        \includegraphics[width=\worldmapwidth]{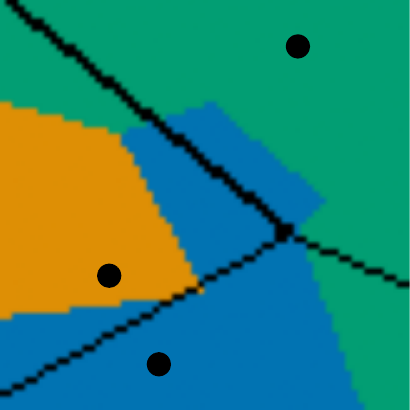}
    }
    \subcaptionbox{Decision Tree\label{subfig:overlayed-boundaries-tree}}{
        \includegraphics[width=\worldmapwidth]{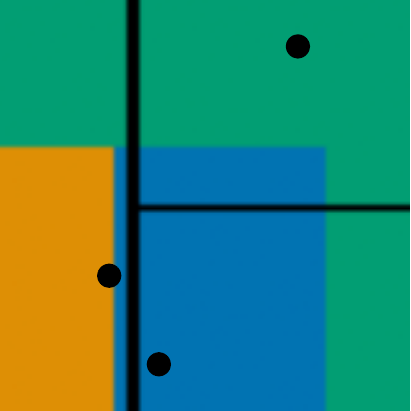}
    }
    \caption{Boundaries (black lines) according to taught classifiers superimposed over the world maps from \cref{fig:ground-truth}.
    We indicate positions of the training samples with black circles -- depending on the experimental condition, these are shown to users, although smaller than depicted here (\cf~\cref{fig:complete-teaching}).}
    \label{fig:overlayed-boundaries}
\end{figure}

The user must be able to see the world map to be able to make an informed decision about which training sample to select next;
at the same time, the taught classifier's current state needs to be communicated as well.
We achieve both by overlaying the decision boundaries according to the current taught classifier over the world map.
See \cref{fig:overlayed-boundaries} for examples of taught boundaries superimposed over world maps.

\subsection{Teaching}\label{subsec:teaching}
Confronted with a world map, we let a user indicate which point to provide as a sample to the learning algorithm by clicking inside the image.
We take that point's coordinates together with its true label (as given in the world map) and add it to the training set.
We immediately (re-)train the taught classifier and draw the resulting decision boundaries on top of the world map.
In addition, the user is shown the current accuracy of the taught classifier over all 64 by 64 pixels of the world map in percent.

When drawn, the taught points can not only function as a reminder
but help the user visually determine a relationship between the taught points and the resulting decision boundaries.
We explicitly investigate the effect of visualizing these points and introduce two conditions for each algorithm.

\begin{figure*}
    \centering
    \subcaptionbox{Nearest Neighbor\label{subfig:complete-teaching-nn}}{
        \includegraphics{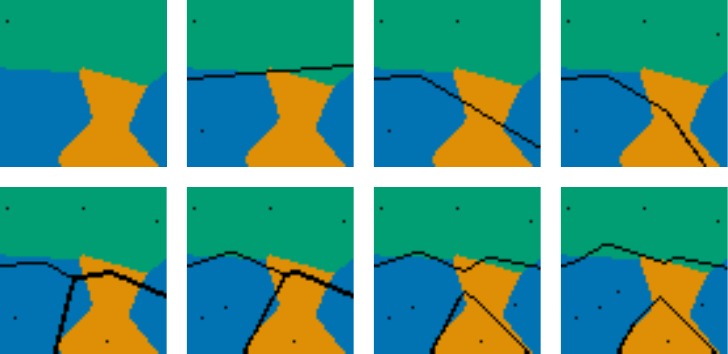}
    }
    \hspace{1cm}
    \subcaptionbox{Decision Tree\label{subfig:complete-teaching-tree}}{
        \includegraphics{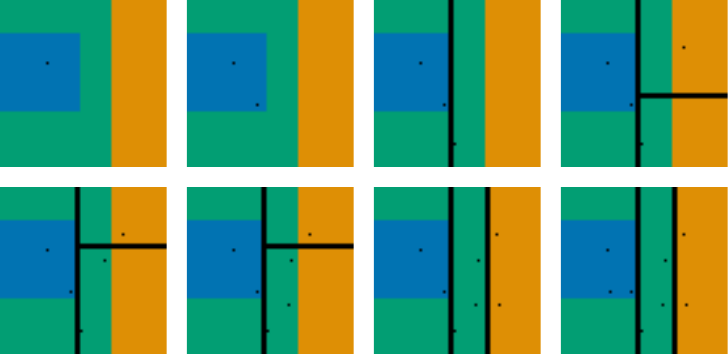}
    }
    \caption{Individual steps of an exemplary teaching process for both the nearest neighbor and decision tree algorithm visualized \emph{with points}.
    }
    \label{fig:complete-teaching}
\end{figure*}

We illustrate the individual steps of an exemplary teaching process in \cref{fig:complete-teaching}.

\subsection{Participant Recruitment}
We recruit a total of \num{800}~participants from Amazon Mechanical Turk\footnote{\url{https://www.mturk.com/}} (AMT)
divided into four conditions with \num{200}~participants each.
We will henceforth refer to the conditions as \nnwith, \nnwithout, \dtwith, and \dtwithout,
depending on the algorithm (nearest neighbor or decision tree)
and whether or not taught points are visualized. %(\cf \cref{table:conditions}).
%\begin{table}
%\centering
%\caption{The four conditions in our study. For each of the two algorithms nearest neighbor and decision tree, we use a visualization with and one without taught points.}
%\label{table:conditions}
%\begin{tabular}{ccc}
%\toprule
%                 & with points & without points \\
%\midrule
%Nearest Neighbor & \nnwith     & \nnwithout \\
%Decision Tree    & \dtwith     & \dtwithout \\
%\bottomrule
%\end{tabular}
%\end{table}

For each participant, we record the order in which the \num{12}~world maps were displayed to them,
the coordinates and order of the points clicked,
the time when they click, and the time our system takes to respond with an update.
To allocate bonus payouts and ensure non-returning participants, we store their AMT identifier.
Beyond this, we neither collect nor store any information -- specifically no personal information.
% <img id="example" src="image.png">
\subsection{Experimental Procedure}\label{subsec:experimental-procedure}
\begin{lstfloat}
\begin{lstlisting}[language=html,breaklines=true,breakindent=0pt,basicstyle=\small\ttfamily]
When you click in an image, that point and its color are sent to the computer. As soon as you click a different color, the computer can guess where color boundaries lie, based on where you clicked. These guesses are shown to you as black lines.

Your goal is for the black lines to align with the actual boundaries between colors as best as possible. The score (100 is best) is shown after each click. Beware, though - the number of clicks is limited for each image.
\end{lstlisting}
\caption{Adjusted excerpt from the introductory text shown to participants.}
%Contains basic HTML to include an image.}
\label{lst:introduction}
\end{lstfloat}
After accepting the task on AMT, a participant is forwarded to our web server and receives HTML with JavaScript.
They are greeted with written instructions and provide their AMT identifiers.
The instructions are identical for each condition, except for an example world map.
Instructions include information about the goal of the teaching task,
how to communicate, and how to interpret the visualization.
See \cref{lst:introduction} for details;
we are careful to provide enough information such that participants can start interacting with the system immediately,
without leaking any algorithm-specific information or biases.
The information contains no indication of how the machine learns to not violate the \emph{lack of explanations} as laid out in \cref{subsec:lack-of-explanations}.

After starting their run, the participant encounters the first warmup world map.
In addition to a world map, we display remaining clicks and the current model's accuracy as the score.
This accuracy is calculated over all 64 by 64 pixels in the world map.
Although our server is fast enough to retrain and render the next image without any noticeable delay,
we enforce at least half a second passing before we accept the next click.
This prevents spammers from rushing the experiment.
After eight clicks, the participant proceeds to the next world map by clicking a button.

On average, our participants need approximately ten minutes to complete the warmup and the twelve actual world maps.
After the final world map, each participant receives a unique code that proves their participation in AMT.
Beyond a base pay, a participant receives a bonus for high scores.
Results from previous studies suggest that such a reward effectively motivated participants to comply with the experimental task~\autocite{bansal2019updates}.

\subsection{Simulated Baseline}\label{subsec:baseline}
Decision trees and nearest neighbors differ in generalization performance in our scenario. Therefore, we determine their baseline performance by simulating participation in the form of randomly selected training points; \num{200}~times for each of the four experimental conditions.

\subsection{Design and Statistical Evaluation}\label{subsec:statistical-evaluation}
We use our data to test whether there are intuitive ML algorithms (\cref{hypothesis1}),
whether different algorithms differ with respect to how intuitive they are (\cref{hypothesis2}),
and evaluate the impact of providing different visualizations (\cref{hypothesis3}).

\subsubsection{User performance \vs simulated baseline data}
We expect to see significant improvement over time for an algorithm to be called intuitive (\cf \cref{subsec:intuitiveness}).
Specifically, we expect to observe better-than-random performance.

\subsubsection{User performance when teaching a nearest neighbor classifier \vs a decision tree classifier}
If different ML algorithms differ in terms of intuitiveness (\cref{hypothesis2}),
we expect significant differences in increasing performance when directly comparing between algorithms.
Thus, we perform a second analysis to compare participants interacting with the nearest neighbor model with participants interacting with a decision tree model.
We adjust for different baseline performances.

\subsubsection{User performance \emph{with point} \vs \emph{without points}}
To assess \cref{hypothesis3}, we evaluate the performance of participants presented with different visualizations.
Concretely, we compare conditions \emph{with points} \vs \emph{without points} for both algorithms.

\subsubsection{Statistical evaluation}
We compare performance over time between users and simulated data \cref{hypothesis1},
between users interacting with different algorithms \cref{hypothesis2}, and different visualizations \cref{hypothesis3} using \emph{R 3.6.2}\footnote{\url{https://www.r-project.org/}} by running separate two-way mixed analyses of variance (ANOVA).
Before each analysis, we check and correct for outliers. Thus, values above the third quartile + 1.5 x the interquartile range or below the first quartile – 1.5 x the interquartile range were considered outliers and removed from further analysis. Further, we assessed compliance with the ANOVA assumption of normality by visually inspecting the correlation between data and the normal distribution in \emph{normal QQ plots}.
In case of heteroscadicity, we apply Greenhouse-Geisser sphericity correction~\autocite{Greenhouse1959MethodsAnalysis}.
Significant interactions or main effects are followed up in post hoc analyses by running simple pairwise comparisons.
We correct of these post hoc analyses for multiple comparisons,
and only report adjusted \(\p\)-values.

\section{Results}\label{sec:results}
\subsection{User Performance \vs Simulated Baseline}
\begin{figure}
    \centering
    \includegraphics[width=(\textwidth-\columnsep)/2]{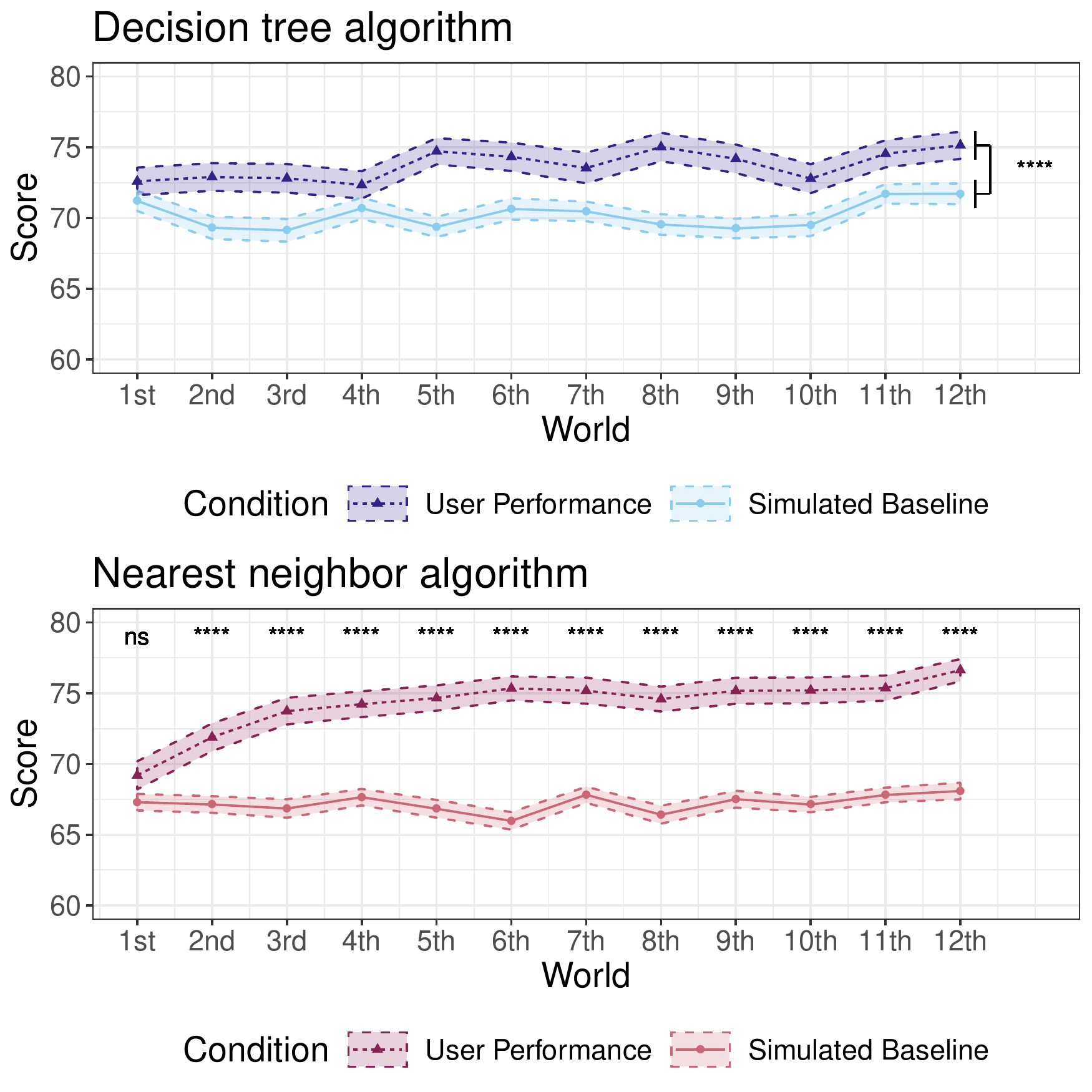}
    \caption{Development of mean scores over worlds for the conditions \dtwith (top) and \nnwith (bottom).
    The central lines of each curve indicate the mean score of the respective condition, and the shaded area around the lines show the range of the standard error of the mean;
    \emph{ns} is not significant; **** indicates \(\p < 0.0001\).}
    \label{fig:results-1}
\end{figure}

In \cref{hypothesis1} we formulate the expectation that intuitive ML algorithms exist,
in the sense that a user interacting collaboratively with such an algorithm gets better over time with regards to the shared objective,
even in the absence of explicit training or explanations.
To statistically assess this hypothesis, we compared data from participants of our study against a random baseline.
In \cref{fig:results-1} we show the development of user performance and simulated over time.
In each case, the data indeed suggests that performance of human users surpasses that of the random baseline.
With the decision tree, participants consistently perform better than the random baseline.
This effect is confirmed by the corresponding two-way mixed ANOVA,
revealing a significant main effect of the factor \emph{condition} (\(F(1, 347) = 55.613\), \(\p < 0.0001\), \(\eta_{\p}^{2} = 0.138\)).
However, there is no main effect over the worlds (\(F(11, 3817) = 1.175\), \(\p = 0.299\), \(\eta_{\p}^{2} = 0.003\)),
nor a significant interaction between \emph{condition} and \emph{world} (\(F(11, 3817) = 0.858\), \(\p = 0.581\), \(\eta_{\p}^{2} = 0.002\)).
Thus, the data does not support the notion of increasing performance with practice for the decision tree compared to a random baseline.

Participants show a marked improvement over time in performance when teaching a nearest neighbor model.
This observation is statistically confirmed by a significant interaction between the factors \emph{condition} and \emph{world} (\(F(10, 3640) = 5.000\), \(\p < 0.0001\), \(\eta_{\p}^{2} = 0.014\)),
indicating that the performances develop differently.
Post hoc analyses reveal significant differences between user performance and random baseline for all worlds but the very first,
highlighting rapidly increasing performance of human participants during interaction.

In conclusion, the data suggests that users interacting with our system for the very first time show performance surpassing that of a random baseline.
Importantly, though, only with the nearest neighbor algorithm does performance increase significantly over time.
Thus, the nearest neighbor algorithm can be considered to be an intuitive ML algorithm, supporting our first hypothesis.
Users interacting with the decision tree algorithm, in contrast,
outperformed the simulated data throughout the experiment, not improving over time.
In order to examine \cref{hypothesis1} more closely,
we set up a statistical model contrasting user performance for both algorithms directly.

\subsection{User performance with a nearest neighbor classifier \vs a decision tree classifier}
We perform an analysis to directly compare performance with the two algorithms.

\begin{figure}
    \centering
    \includegraphics[width=(\textwidth-\columnsep)/2]{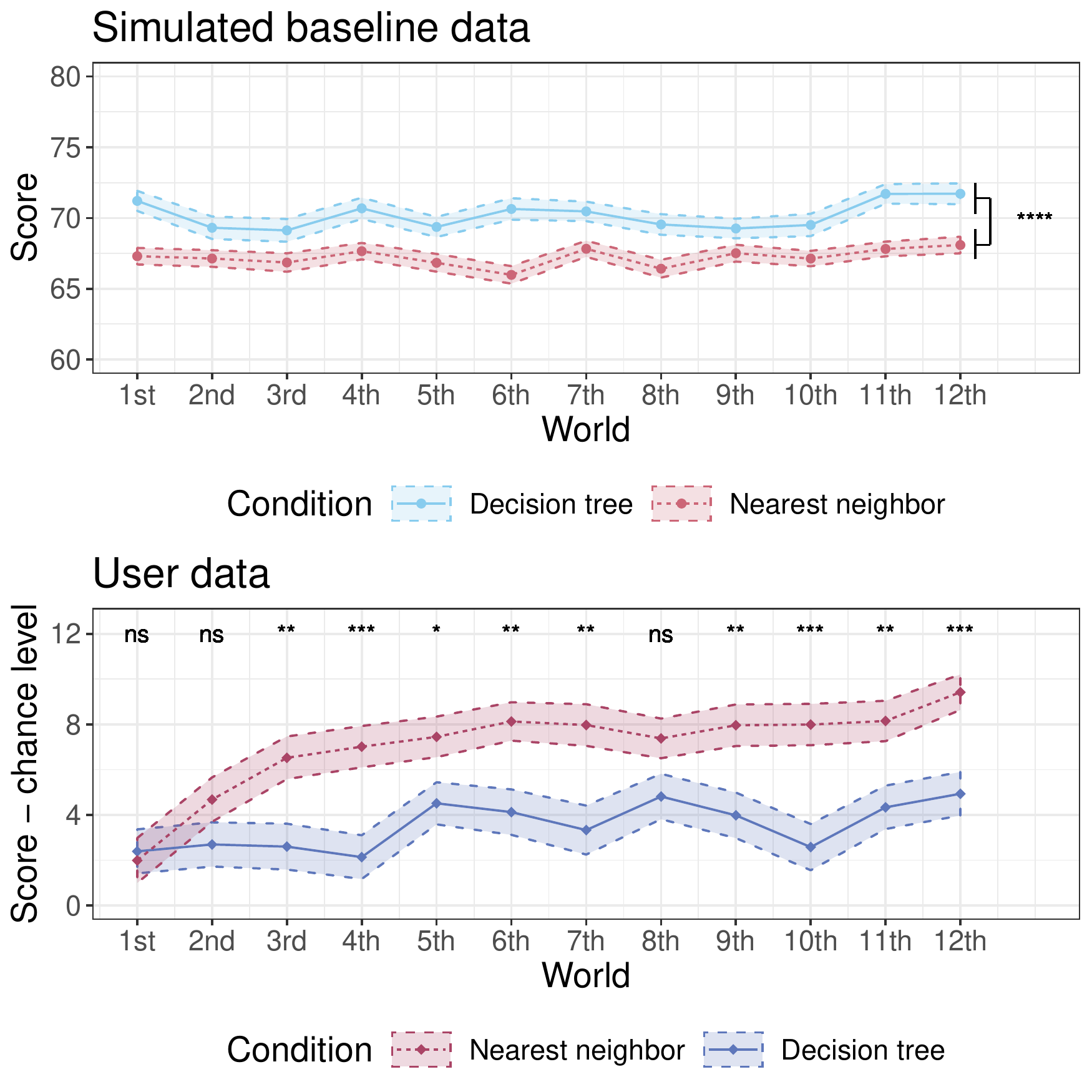}
    \caption{Development of mean scores over worlds for the simulated baseline data (top) and the normalized behavioral scores (bottom). User scores are normalized by subtracting the respective mean value across worlds of the random baselines. The central lines of each curve indicate the mean score of the respective condition, and the shaded area around the lines show the range of the standard error of the mean;
    \emph{ns} is not significant; * indicates \(\p < 0.05\); ** indicates \(\p < 0.01\); *** indicates \(\p < 0.001\); **** indicates \(\p < 0.0001\).}
    \label{fig:results-2}
\end{figure}

Before the actual comparison, note an important observation from the previous analysis:
mere visual inspection suggests a difference in random baselines between the two algorithms (\cf~\cref{fig:results-2}, top).
Indeed, this suspicion is confirmed by a significant main effect of factor \emph{condition} in a corresponding two-way mixed ANOVA comparing the random baselines directly (\(F(1, 351) = 139.563\), \(\p < 0.0001\), \(\eta_{\p}^{2} = 0.284\)).
Thus, to counteract a possibly confounding effect driven by the difference in random baselines,
we normalize scores by subtracting the respective mean value across worlds of the random baselines prior to our statistical analysis.
In the lower plot in \cref{fig:results-2} we show how for both groups,
normalized user scores consistently increase over worlds, with a steeper increase for the nearest neighbor algorithm.
Participants achieve higher normalized scores with the nearest neighbor algorithm than with a decision tree.
This pattern is statistically confirmed in the corresponding two-way mixed ANOVA,
with a significant interaction of the factors \emph{condition} and \emph{world} (\(F(10, 3598) = 2.033\), \(\p = 0.026\), \(\eta_{\p}^{2} = 0.006\)).
Additionally, we detect significant main effects of \emph{condition} (\(F(1, 358) = 18.471\), \(\p < 0.0001\), \(\eta_{\p}^{2} = 0.049\)) and \emph{world} (\(F(10, 3598) = 4.804\), \(\p < 0.0001\), \(\eta_{\p}^{2} = 0.013\)).
Post hoc analyses reveal that the interaction effect is driven by significantly improved performance starting from the third world for the nearest neighbor algorithm,
compared to the decision tree algorithm.
Thus, we may conclude that participants improve at a higher rate with the nearest neighbor algorithm.
Consequently, following our definition of intuitiveness,
this pattern indeed supports the notion that the nearest neighbor algorithm is more intuitive than the decision tree algorithm.

\subsection{User performance \emph{with points} \vs \emph{without points}}

\begin{figure}
    \centering
    \includegraphics[width=(\textwidth-\columnsep)/2]{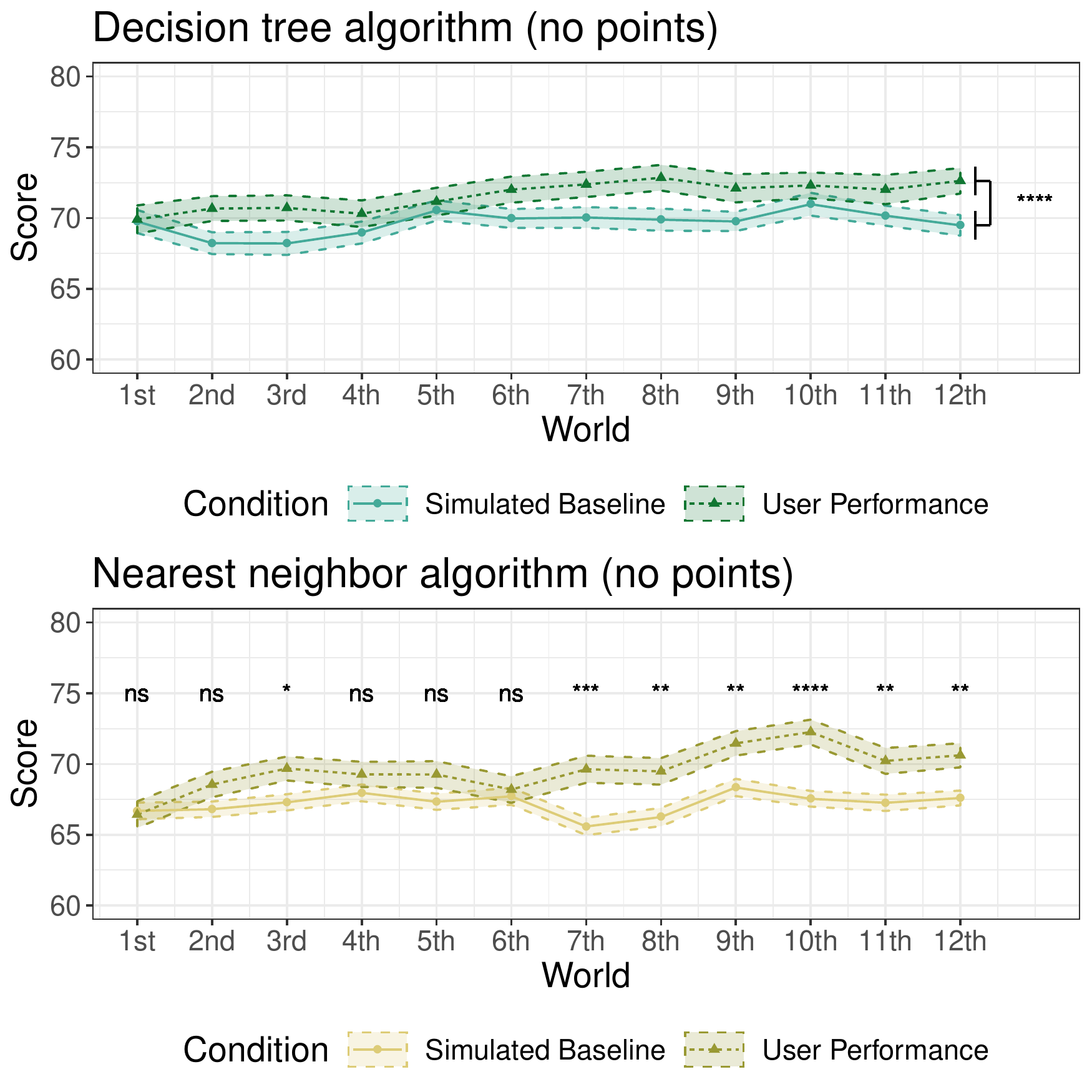}
    \caption{Development of mean scores over worlds for the conditions \dtwithout and \nnwithout.
    The central lines of each curve indicate the mean score of the respective condition, and the shaded area around the lines show the range of the standard error of the mean;
    \emph{ns} is not significant; * indicates \(\p < 0.05\); ** indicates \(\p < 0.01\); *** indicates \(\p < 0.001\).}
    \label{fig:results-3}
\end{figure}

In \cref{hypothesis3}, we posit that -- even without external explanations --
different visualizations related to the current state may influence the rate of improvement.
To explore this supposition, we acquired additional data without displaying the taught points.
Comparing this new data with the respective random baselines yields a pattern remarkably similar to the condition \emph{with points} (see \cref{fig:results-3}).
For the decision tree algorithm, even \emph{without points},
users again show above-chance performance across all worlds.
This is confirmed by a significant main effect of \emph{condition} (\(F(1, 354) = 41.298\), \(\p < 0.0001\), \(\eta_{\p}^{2} = 0.104\)) in the corresponding two-way mixed ANOVA.
Additionally, unlike \emph{with points},
there is a significant main effect of factor \emph{world} (\(F(11, 3894) = 2.185\), \(\p = 0.013\), \(\eta_{\p}^{2} = 0.006\)).
This effect is potentially driven by lower scores for users during the first five worlds.

For the nearest neighbor algorithm, users in \nnwithout replicate the exact same pattern as in \nnwith: performance improves visibly (see \cref{fig:results-3}, bottom).
This is confirmed by a significant interaction between factors \emph{condition} and \emph{world} (\(F(10, 3863) = 2.44\), \(\p = 0.006\), \(\eta_{\p}^{2} = 0.006\)).
Closer inspection by pairwise post hoc comparisons shows that this effect is driven by significant differences between the random baseline and participants from world seven onward.
In contrast, when users interacting with this algorithm have information about previously selected points available,
these differences are detectable from world two onward (see \cref{fig:results-1}, bottom).
This may be taken as first evidence in favor of \cref{hypothesis3}.

\begin{figure}
    \centering
    \includegraphics[width=(\textwidth-\columnsep)/2]{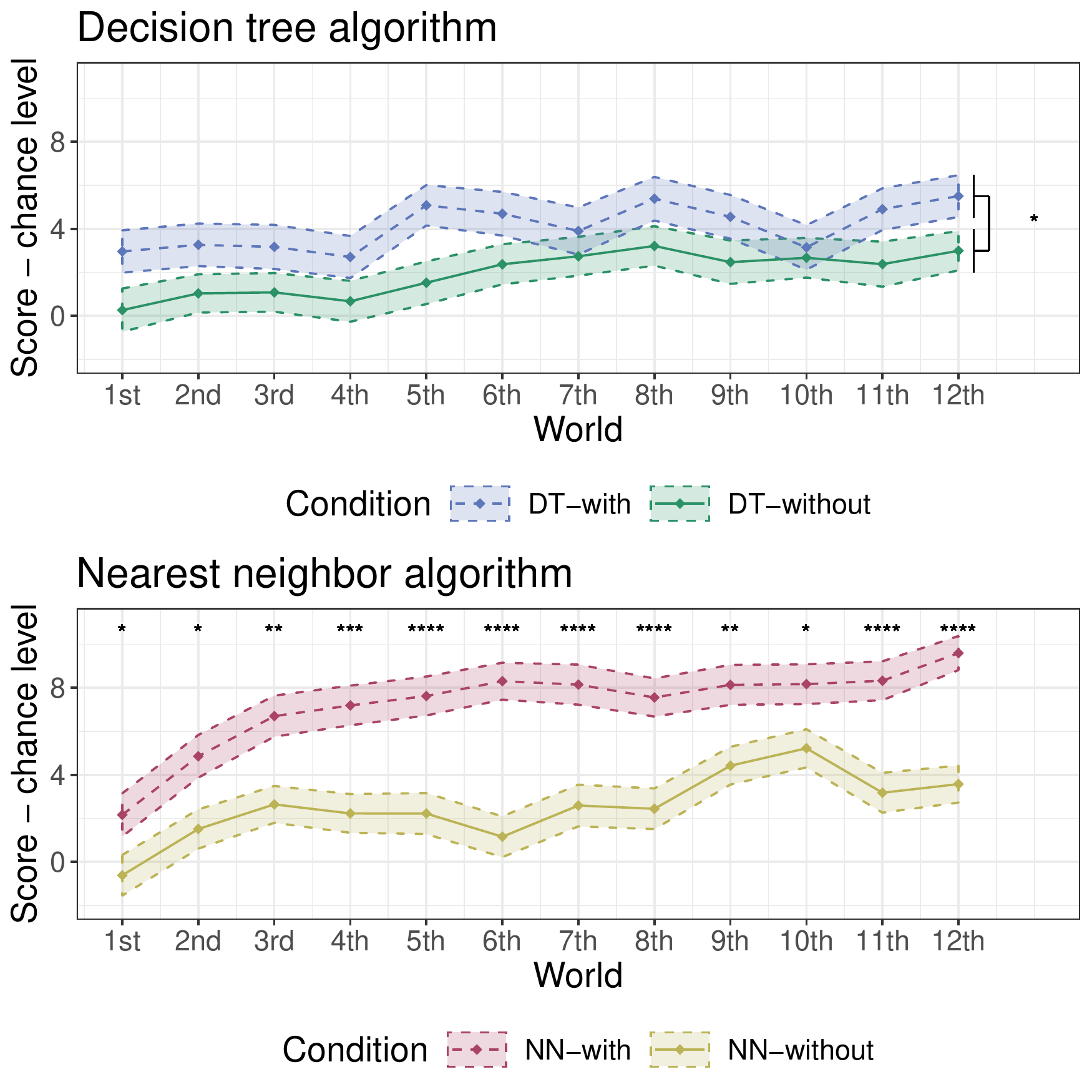}
    \caption{Development of mean normalized scores over worlds for the conditions \dtwith \vs \dtwithout (top) and \nnwith \vs \nnwithout (bottom).
    Data is normalized by subtracting the mean value of the random baselines.
    The central lines of each curve indicate the mean score of the respective condition, and the shaded area around the lines show the range of the standard error of the mean;
    * indicates \(\p <0.05\); ** indicates \(\p < 0.01\); *** indicates \(\p < 0.001\); **** indicates \(\p < 0.0001\).}
    \label{fig:results-4}
\end{figure}

We directly compare normalized scores derived from users interacting with the nearest neighbor algorithm and the decision tree algorithm in conditions \emph{with points} and \emph{no points}, respectively.
Level and trajectory of random baselines does not differ between the conditions \emph{with points} and \emph{with points} for the decision tree (no significant main effect of \emph{group};
\(F(1, 398) = 1.248\), \(\p = 0.265\), \(\eta_{\p}^{2} = 0.003\)) and nearest neighbor (no significant main effect of \emph{group}; \(F(1, 398) = 0.025\), \(\p = 0.874\), \(\eta_{\p}^{2} < 0.0001\)) algorithms, respectively.
Therefore, we normalize the data by the mean value of the random baselines across the conditions \emph{with points} and \emph{without points} (see \cref{fig:results-4}).
We see that performance \emph{with points} is generally better than \emph{without points}.
This observation is significant, shown by main effects of factor \emph{group} in the corresponding two-way mixed ANOVAs (Decision tree: \(F(1, 350) = 6.061\), \(\p = 0.014\), \(\eta_{\p}^{2} = 0.017\);
Nearest neighbor: \(F(1, 369) = 36.449\), \(\p < 0.0001\) ,\(\eta_{\p}^{2} = 0.09\)).
For the nearest neighbor algorithm, there is a significant main effect of factor \emph{world} (\(F(9, 3432) = 11.008\), \(\p < 0.0001\), \(\eta_{\p}^{2} = 0.029\)).
The interaction of \emph{group} and \emph{world} is significant for the nearest neighbor algorithm as well (\(F(9, 3432) = 1.871\), \(\p = 0.049\), \(\eta_{\p}^{2} = 0.005\)).
Post hoc analyses reveal significant differences between \nnwith and \nnwithout that get increasingly more pronounced over time:
teaching a nearest neighbor model, participants \emph{with points} increase in performance more quickly than participants \emph{without points}.
Interestingly, this is only the case for the nearest neighbor algorithm which, based on our previous analyses, appears more intuitive than the decision tree algorithm.
Overall, the main effects of \emph{group} observed for each algorithm support \cref{hypothesis3}.
Additionally, a more informative visualization may be particularly beneficial for an algorithm that is per se intuitive.

Note that training points directly correspond to the internal representation of a nearest neighbors classifier,
while decision boundaries directly correspond to a decision tree's internal representation.

\section{Discussion}\label{sec:discussion}
In this work, we propose intuitiveness as a core property of ML algorithms that strongly impacts how easy it is for users to interact with an algorithm without any explicit instruction or training.
We corroborate the usefulness of this approach in a large-scale human-grounded~\autocite{Doshi2017RigorousInterpretable} evaluation, assessing collaborative user-machine interaction in an active teaching study in which naive users select training samples to efficiently teach a ML algorithm.

Through the experimental design of our study, we focus on two basic, well established, and highly relevant ML algorithms that can learn from few data points.
Additionally, we successfully formalize a straight-forward and immediate visualization of teaching progress.
Despite its simplicity, this scenario could appear in its current form when teaching spatial semantic concepts to a household assistant, which underlines is applicability and relevance.

Our results reveal a range of valuable insights with important implications for human-machine interaction.

First, we demonstrate that random teaching provides a constant baseline, against which improvements can be reliably measured.
Based on these comparisons against chance, we verify that both algorithms allow better-than-random teaching.
If selecting points at random consistently generated scores comparable to those of participants, this would suggest that participants either did not understand the task, were overwhelmed by it, or did intentionally not comply with the experimental instructions. In our study, however, naive users surpass the performance of a random baseline.

Second, we show that some ML algorithms are more intuitive than others.
Specifically, the nearest neighbor algorithm appears more intuitive than the decision tree algorithm when taught by a naive human user.

Third, our study reveals that different visualizations of the teaching progress affect performance:
generally, better scores are achieved when more information is provided.
We discover that this added information may be particularly beneficial when the taught algorithm is intuitive, indicated by the increase in user performance for the nearest neighbor (\nnwith and \nnwithout), but not the decision tree (\dtwith and \dtwithout).
This difference is astonishing, given that one and the same experimental manipulation -- visualizing or not visualizing previously selected training samples -- has a differential effect of how well users interact with and improve the system.
A possible explanation for this contrast may be that for the nearest neighbor algorithm, the points directly correspond to the model, while the link between points and the collection of if-then-else rules in a decision tree is less direct.

\subsection{Limitations \& Future Work}\label{subsec:limitations-future-work}
Despite the large scale of our user study --
comprising data from \num{800}~participants interacting with our system --
we cannot guarantee an unbiased representation of behavioral patterns present in the general population.
Specifically, our current analysis did not warrant assessing demographic information, individual personality or cognitive traits in each participant in detail.
Thus, we were not able to identify factors that may affect user performance on an inter-individual level.
Recent work in the domain of interpretable ML suggests that individual differences in the ability to understand and apply the results of automated ML systems may depend on individual personality traits,
such as those associated with mathematical ability and gist processing~\autocite{Gleaves2020IndividualUserDifferences}.
A similar evaluation with respect to inter-individual differences in intuitiveness as a property of algorithms remains an important research gap to be addressed.

Our in-depth analysis of the intuitiveness of two concrete algorithms does not include an evaluation of what characteristics make an algorithm more or less intuitive.
While the reported results indicate differences in terms of the mental burden imposed on the user, the scope of this work only allows limited insights what factors determine cognitive complexity of algorithms. 
We demonstrate that the amount of information displayed eases the effort required to identify efficient teaching strategies. 
Identification of further factors impacting cognitive complexity of ML-algorithms is an important avenue for future studies.
Moreover, there is still room for comprehensive examinations of further ML algorithms,
such as Support Vector Machines or Artificial Neural Networks.
Such prospective work may lead to general principles and ultimately guide the design of future intuitive human-machine interfaces.

The task participants face in our active teaching study is a contrived abstraction of a real-world human-machine interaction.
During the task, achieving low scores has no real negative consequence for the user,
which is why we use monetary incentive to motivate.
Extending the task and design to project greater perceived real-world impact constitutes a challenging
-- though certainly worthwhile -- avenue for future work,
considering the importance of intrinsic motivation in quality-focused tasks~\autocite{Cerasoli2014IntrinsicMotivation}.

Despite these limitations, our results show very promising first steps towards an experimental evaluation of the intuitiveness of teachable algorithms.
Insights into users' internal representations~\autocite{Hindemith2020TechnicalRobots} of the machine might lead to an understanding of which characteristics of algorithms support intuitiveness.

\subsection{Conclusion}\label{subsec:conclusion}
We propose intuitiveness as a core property of ML algorithms, largely impacting how easy it is for users to interact with an algorithm without any explicit instruction or training.
In real-world interpersonal interactions and decision making, action is often taken based on \emph{gut feeling} or intuition, rather than conscious analytical reasoning.
Users interacting with artificial systems tend to automatically and subconsciously transfer behavioral patterns from interpersonal experiences to this new setting~\autocite{Nass1994ComputersSocialActors},
so it seems plausible that users also rely on intuition in human-machine interaction.
Our work provides initial evidence for this notion.
Specifically, in our human-grounded evaluation we demonstrate that two well-established ML algorithms indeed vary with respect to how intuitive they are,
and that different visualizations influence how quickly users improve in teaching these algorithms.
This work is a first step to systematically evaluate the efficacy of human-machine interactions with a particular focus on the impact of intuitiveness, to be extended by future evaluations including a wider range of more complex algorithms and higher dimensional data.
Thus, our framework of intuitiveness in active teaching may ultimately guide the design of future human-machine interfaces that are intuitively accessible.

%%%%
% References
%%%%
\AtNextBibliography{\raggedright\small}
\printbibliography
\end{document}